\documentclass[iop]{emulateapj}

\usepackage{hyperref}

\begin{document}


\title{At a Crossroads: Stellar Streams in the South Galactic Cap}

\author{Carl J. Grillmair}
\affil{IPAC, Mail Code 314-6, Caltech, 1200 E. California Blvd., Pasadena, CA 91125}
\email{carl@ipac.caltech.edu}

\begin{abstract}

We examine the distribution of old, metal-poor stars in a portion of
the recently released PanSTARRs survey. We find an interesting
confluence of four new cold stellar stream candidates that appear to
converge on or pass near the south Galactic pole. The stream
candidates, which we designate Murrumbidgee, Molonglo, Orinoco, and
Kwando, lie at a distance of $\approx 20$ kpc and range in length from
$13\arcdeg$ to $95\arcdeg$, or about 5 to 33 kpc. The stream
candidates are between 100 and 300 pc in width, and are estimated to
contain between 3000 and 8000 stars each, suggesting progenitors
similar to modern day globular clusters. The trajectories of the
streams imply orbits that range from hyperbolic to nearly
circular. The Molonglo stream is nearly parallel to, at the same
distance as, and offset by only $2.5\arcdeg$ from the previously
discovered ATLAS stream, suggesting a possible common origin. Orinoco
and Kwando also have similarly shaped, moderately eccentric, obliquely
viewed orbits that suggest distinct progenitors within a common,
larger parent body.

\end{abstract}


\keywords{Galaxy: Structure --- Galaxy: Halo}

\section{Introduction}

With the advent of wide-field digital sky surveys, recent years have
seen the discovery of several dozen highly collimated, cold stellar
debris streams in the Galactic halo (see \citet{grillmair2016} and
\citet{smith2016} for reviews). These streams are clear evidence of
the build-up of the halo through the accretion or destruction of
ancient star clusters and dwarf galaxies. Moreover, these streams are
providing us with sensitive new probes of the Galactic potential and
the distribution of dark matter. Analyses of the positions and motions
of stars in these streams enable us to constrain both the local and
global shape of the potential (e.g. \citet{kupper2015, bovy2016}), while
the lengths, asymmetries, and discontinuities in the streams can be
used to infer the properties of other Galactic components
(e.g. Pearson et al 2017) and of the dark matter itself
\citep{carlberg2009, yoon2011, carlberg2013}.

Gaia is expected to greatly increase our knowledge of streams and
substructures in the halo (Gaia Collaboration et al.(2016)). However,
Gaia's magnitude cutoff may limit its ability to independently detect
the less populous and more far-flung streams. Knowing in advance the
locations and trajectories of more distant streams will enable us to
both untangle what will no doubt be a very complex dataset, and to
identify and characterize tracer stars on the red giant or horizontal
branches. Accurate proper motions for these tracers would greatly
enhance the utility and constraining power of these streams for 
models of the Galaxy.

While most known streams have been discovered in the Sloan Digital Sky
Survey \citep{abazagian2009}, the recent public release of the
Pan-STARRs catalog (PS-1, \citet{chambers2016}) enables us to extend
our searches over a significantly larger area of
sky. \citet{bernard2016} recently carried out a survey of PS-1 and, using
a matched filter to count stars with colors and magnitudes consistent
with an old, metal poor population, discovered five new stellar debris
streams. In this paper we use a similar technique to more closely
examine the region in the south Galactic cap. We describe our
detection method in Section \ref{analysis}. We discuss each of our
detections in more detail in subsections \ref{molonglo} through
\ref{kwando}. Concluding remarks are given in Section
\ref{conclusions}.

\section{Analysis} \label{analysis}

We make use of the photometric catalog from data release 1 of
PanSTARRs (Chambers et al. 2016). After some experimentation
with photometry from the MeanObject and ForcedMeanOject tables, we
elected to proceed using the MeanObject table. While this table is
less deep than the forced photometry table, we find that in this
application the significantly smaller photometric uncertainties in the
MeanObject table more than compensate for the somewhat reduced depth.
From our experience with Sloan photometry, we elected to download and
analyze only the $g, r,$ and $i$ measurements, which are best suited
for distinguishing among the relatively blue stars at the main
sequence turn-off and below.

To ensure that our sample consisted primarily of stars (as opposed to
galaxies), we selected all sources with

\begin{eqnarray}
i_{PSF} - i_{Kron} < -0.307 + 0.0442 i_{PSF} \nonumber \\
- 0.00777 i_{PSF}^2 + 0.000113 i_{PSF}^3
\end{eqnarray}

\noindent This is slightly more involved than the simple $i_{PSF} -
i_{Kron} < 0.05$ cut suggested by \citet{chambers2016} but better
reflects the locus of stars at faint magnitudes. After some
experimentation, we elected to use all stars with $g <
21.8$. Photometry was dereddened using the DIRBE/IRAS dust maps of
\citet{schleg98}, corrected using the prescription of
\citet{schlafly2011}.

We constructed matched filters as described by \cite{rockosi2002} and
\citet{grillmair2009} using the $g - r$ and $g - i$ distributions of
stars in the globular cluster NGC 5053.  This cluster has [Fe/H]
$\approx -2.29$ \citep{harris1996}, yielding filters designed to
highlight very low metallicity populations.  The luminosity function
is modeled as a simple powerlaw, with $N \propto g^{0.3}$. This gives
somewhat greater weight to low mass stars than would a peaked,
globular cluster luminosity function, but it accords with our
expectations concerning mass segregation in clusters, is consistent
with observations of Pal 5's tidal tails \citep{koch2004}), and
appears to work quite well.  The foreground population was sampled
over $\approx 50$\% of the area in the south Galactic cap, avoiding
regions occupied by the Sagittarius and Triangulum-Pisces streams
\citep{bonaca2012} as well as various globular clusters in the
field. The signal/foreground probabilities computed in each color were
then multiplied together before summing the probabilities by position
on the sky. We applied the filters to stars in the south Galactic cap
after successively shifting the filters in 0.1 magnitude intervals to
distance equivalents ranging from one to 100 kpc.

Figure 1 shows the result of applying our matched filter at a distance
of 20 kpc. Five streams are apparent, all passing within a few degrees
of the globular cluster NGC 288 and the south Galactic pole (SGP). The
ATLAS stream, first discovered by \citet{koposov2014} and subsequently
extended by \citet{bernard2016} is easily seen, ending at $\delta \sim
-15\arcdeg$. We see no further extensions of ATLAS along the
best-fitting great circle determined by \citet{bernard2016}.

Some of the new features in Figure 1 are only barely perceptible, if
at all, in Figures 1 and 2 of \citet{bernard2016}. We find that
different projections can have a very significant effect on the
detectability of very tenuous features, and the streams in Figure 1
(including ATLAS) are much less prominent in either equatorial or
Galactic Mercator or Aitoff coordinate projections. We also note that
the features in Figure 1 (including the ATLAS stream) largely fade
from detectability using a matched filter shifted more than $\pm 0.3$
magnitudes from the correct distance modulus for the stream.
Moreover, the 20 kpc distance of the features in Figure 1 falls on the
borderline between the intermediate and distant maps generated by
\citet{bernard2016}, in effect reducing signal-to-noise ratio at this
particular distance, as compared with distances 0.3 magnitudes closer
or further away.

Figure 2 shows the distribution of {\it E(B-V)} over the same region of
sky as in Figure 1 \citep{schleg98}. While we apply a magnitude cut to
the sample {\it after} applying dereddening corrections, inaccuracies in the reddening model could pull in either too few or too many stars from
redder and fainter regions of the color magnitude
distribution. However, Figure 2 shows no significant features that can
be matched to the lengths and positions of our stream candidates. We
conclude that they are not artifacts of reddening-induced
completeness variations. We designate the stream candidates as Molonglo,
Murrumbidgee, Orinoco, and Kwando. We discuss each of these in turn below.

\subsection{Molonglo} \label{molonglo}

The Molonglo stream, detected at the $\approx 10\sigma$ level, is some
$22\arcdeg$ long, extending from near the globular cluster NGC 288
across both the bright and faint branches of the Sagittarius
stream. While pointing almost directly at NGC 288, the stream is at
more than twice the 8.8 kpc distance of the cluster
\citep{harris1996}. Moreover, the measured proper motion of NGC 288
\citep{dinescu1997} would predict a stream oriented more or less
east-west, along the southern limit of the PS1 survey. Molonglo is
therefore unlikely to be related to NGC 288.

In equatorial coordinates, the trajectory of Molonglo can be modeled
to $\sigma = 0.1\arcdeg$ using:

\begin{eqnarray}
\alpha = 345.017 - 0.5843 \delta + 0.0182 \delta^2 
\end{eqnarray}

\noindent over the stream extent of $-24.5\arcdeg < \delta <
-12\arcdeg$. The full-width-at-half-maximum (FWHM) of the stream is
about 30 arcmin. At a distance of 20 kpc, the length and breadth of
the stream are therefore 7.7 kpc and 170 pc, respectively. The latter
suggests a relatively small or low mass progenitor such as a globular
cluster. A background-subtracted, color-magnitude Hess diagram is
shown in Figure 3. While obviously contaminated by the Sagittarius
stream, Molonglo appears to have a discernible subgiant branch and
lower main sequence. As with other streams, shifting the filter
brightward or faintward by more than half a magnitude causes the
structure to effectively vanish. Estimating the distribution of foreground
stars using a heavily smoothed version of Figure 1 and then simply
counting stars along the stream within the $3\sigma$ color-magnitude
envelope yields an estimated total of $311 \pm 95$ stars to $g =
21.8$. Integrating over a globular cluster-like luminosity function
then predicts a total of $3700 \pm 1100$ stars over the $22\arcdeg$
length of Molonglo. This is certainly comparable to modern day
globular clusters and suggests that Molonglo might 
once have been a globular cluster itself.

While stars in streams do not precisely follow the same orbits, or the
orbits of their progenitors \citep{eyre2011}, the observed
trajectories of streams can put some constraints on their orbit
shapes. These in turn can be used to identify orbit families or
possible progenitors. Starting from a designated fiducial point midway
along the stream, we use a simple model of the Galaxy with a spherical
halo \citep{allen1991} to integrate a range of possible
orbits. Assuming an uncertainty of $0.2\arcdeg$ in the positions of
ten normal points evenly spaced along the stream, we use the $\chi^2$
in the positional agreement to arrive at 90\% confidence limits for the proper
motions and heliocentric velocities at the fiducial point necessary to
match the trajectory of the stream. We then push each of these
quantities to their 90\% limits, integrate the orbits, and choose the
extreme cases to put approximate limits on the orbital
parameters. These limits are given in Table 1. The best-fitting orbits
are shown in Galactic cartesian coordinates in Figure 4.

With no detectable curvature on the sky, the trajectory of Molonglo
evidently requires that the progenitor be essentially unbound from the
Galaxy. This is somewhat surprising, and one might wonder how the
progenitor could have been so extensively disrupted on so benign an orbit. The
progenitor must either have been very loosely bound to begin
with, or it may have been disrupted by an encounter with some other
resident of the Local Group.

\subsection{Murrumbidgee} \label{murrumbidgee}

The Murrumbidgee stream is visible (though barely so at its southern
end) over the entire field of view in Figure 1, from the SGP to within
$5\arcdeg$ of the Galactic plane. Using the T-statistic of
\citet{grillmair2009}, the portion of the stream with $-65\arcdeg < b
< -30\arcdeg$ is detected at approximately the $6\sigma$ level. The
color-magnitude distribution for this portion of the stream is shown
in Figure 3 and shows both a weak subgiant branch, along with a somewhat
stronger lower main sequence. It also suggests that the stream could
be slightly more metal rich than our adopted search filter. On the
other hand, application of a filter based on the color-magnitude
distribution of stars in Pal 5 ([Fe/H] = -1.43) yields significantly
less signal-to-noise ratio, indicating that the off-locus population
at $g \approx 20.6$ in Figure 3 is probably unrelated to the stream.

The trajectory of the stream can be modeled to $\sigma =0.14\arcdeg$
using:

\begin{eqnarray}
\alpha = 367.893 - 0.4647 \delta - 0.00862 \delta^2 + 0.000118 \delta^3 \nonumber \\
+1.2347 \times 10^{-6} \delta^4 - 1.13758 \times 10^{-7} \delta^5
\end{eqnarray}

\noindent The stream is some $95\arcdeg$ long, and some 22 arcmin
across (FWHM). At a distance of 20 kpc, these translate to a length of
33 kpc, and a width of 125 pc. The width is again comparable to those
of known or presumed globular cluster streams such as Pal 5
\citep{odenkirchen2003}, NGC 5466 \citep{belokurov2006,
  grillmair2006a}, or GD-1 \citep{grillmair2006b}. For the most
visible, $45\arcdeg$-long portion of the stream north of the faint
Sagittarius stream, subtracting a smoothed foreground distribution and
counting stars in the color-magnitude envelope yields a total of $310
\pm 120$ stars. Doubling this to account for portions of the stream
heavily contaminated by the disk or by Sagittarius stream stars, and
integrating over a globular cluster-like luminosity function, we
arrive at an estimated total of $7400 \pm 2900$ stars in the visible
stream. This is again consistent with a globular cluster progenitor.

Within the limitations of our adopted Galactic model, an orbit fit to
Murrumbidgee puts reasonably tight constraints on the orbital
plane. The best-fitting orbit is highly inclined and nearly circular,
with perigalacticon occurring just north of the faint Sagittarius
stream. The orbit passes within 2 kpc of Pal 5 and and 3 kpc of NGC
1261. The orbital path of Pal 5 is fairly well described by its tidal
tails, and its orbit pole is inclined by more than $30\arcdeg$ to that
of Murrumbidgee, ruling out any physical association. As we have no
proper motion information for NGC 1261, we cannot say at this point
whether there may be a physical association. If we allow the
uncertainties in the fitted proper motions and radial velocities to
extend to their $1\sigma$ limits, we find that IC 4499 and NGC 5634
also lie with 3 kpc of an orbit integration.

\subsection{Orinoco} \label{orinoco}

The eastern $19\arcdeg$ of Orinoco (the portion nearest to the SGP) is
detected at the $\approx 15\sigma$ level, and is interesting both for
its pronounced curvature and its projected overlap with the ATLAS
stream. A portion of the stream appears to be visible in
\citet{koposov2014}, though not a length sufficient to have been
identified as a separate stream. The western portion of the stream (extending
downwards in Figure 1) is much weaker and somewhat
conjectural. Fitting an orbit to just the eastern $19\arcdeg$ yields a
best-fit trajectory that lies alongside and separated by less than a
degree from the very faint feature in Figure 1 with $-62\arcdeg < b <
-48\arcdeg$. Including this faint feature in the fit results in a no
less plausible orbit. If indeed the two features are part of the same
stream, then it would appear that there is a $\approx 10\arcdeg$ gap
between the two. At this level of significance, it is plausible that
this region is simply below our detection threshold. On the other
hand, it may also constitute a physical gap in the stream, generated
perhaps by a close encounter with a dark matter subhalo or some
other Galactic constituent. Finally, we must recognize that this
western feature could be an unrelated stream, or not a stream at all.

The easternmost $19\arcdeg$ of Orinoco can be modeled to $\sigma =
0.13\arcdeg$ using:

\begin{eqnarray}
\delta = -25.5146 + 0.1672 \alpha - 0.003827 \alpha^2 \nonumber \\
- 0.0002835 \alpha^3 - 5.3133 \times 10^{-6} \alpha^4
\end{eqnarray}

\noindent This section of the stream has a FWHM of about 40 arcmin,
indicating a physical breadth of about 240 pc.  This is somewhat
broader than the streams above, and may be partly a consequence of
confusion with the overlapping ATLAS stream. It may also be indicative
of a somewhat more massive progenitor, or more significant or extended
heating of the stream \citep{carlberg2009}. Estimating foreground
contamination and counting stars in the appropriate color-magnitude
envelope, we find a total of $225 \pm 95$ stars in the eastern
$19\arcdeg$ of the stream.  Integrating over a globular cluster-like
luminosity function, we estimate a total population of $2700 \pm 1100$
stars.

The eastern (upper) portion of Orinoco apparently bends over and
overlaps with a portion of the ATLAS stream
\citep{koposov2014,bernard2016}. However, an orbit fit to Orinoco
suggests only a few degrees of overlap before the stream diverges from
ATLAS in a more southerly direction. We have attempted to fit an orbit
to both Orinoco and the southern portion of ATLAS, but the resulting
$\chi^2$ is much larger than for either stream alone. We conclude that
ATLAS and Orinoco are not different parts of the same stream.

The best-fit orbit for Orinoco passes within 1 kpc of NGC 288, and of NGC
6356. It also passes within 2 kpc of the nearby globulars NGC 6121,
NGC 6284, and NGC 6397. From inspection of Figure 1 we know that
Orinoco cannot be associated with NGC 288. While the stream appears to
arc around the cluster, it does not attach to the cluster in the
characteristic ``S-shape'' we expect (e.g. Pal 5). Using proper motion
measurements for the remaining four clusters \citep{dinescu1999,
  casettidinescu2010, casettidinescu2013}, we find that all of them have
orbits confined to less than 5 kpc vertically from the Galactic
disk. The estimated distance of Orinoco ($20 \pm 3$ kpc), combined
with the best-fit orbit parameters in Table 1, rule out any
association between the stream and these particular clusters.

\subsection{Kwando} \label{kwando}

Faint and not particularly long or continuous, Kwando would not
normally have been identified as a stream candidate were it not for
its similarity and proximity to Orinoco. The same round-the-pole arc,
shifted only a few degrees eastward from Orinoco, was sufficiently
striking that we opted to consider it. The stream is detected at the
$7\sigma$ significance level and, over the range $19\arcdeg < \alpha <
31\arcdeg$, the $13\arcdeg$ arc of Kwando is matched to $0.1\arcdeg$
by a polynomial of the form:

\begin{equation}
\delta = -7.817 - 2.354 \alpha + 0.1202 \alpha^2 - 0.00215 \alpha^3 
\end{equation}

With a FWHM of 22 arcmin, Kwando is evidently only 130 pc across, once
again in the realm of presumed globular cluster streams. A
color-magnitude selected count of stars yields $120 \pm 52$ stars over
the $13\arcdeg$ arc of the stream, or $1400 \pm 600$ when integrated
over a globular cluster-like luminosity function. In terms of linear
density, the $9 \pm 4$ stars per degree actually exceeds the $8 \pm 3$
stars per degree measured for Murrumbidgee.

As might be expected based on
appearance and proximity, the best fit to the orbit of the stream
shows fairly similar parameters to that of Orinoco. To within the
uncertainties, the inclination and orbit poles are identical. This
suggests that, while these streams obviously had distinct progenitors,
those progenitors themselves may have formed in the same cloud, or
orbited a common parent body.

For Kwando, the best-fit orbit comes within 1 kpc of 47 Tuc. Based on
the proper motion measurements of \citet{watkins2017}, an orbit
integration for 47 Tuc reveals that the cluster orbit is confined to
within 5 kpc of the Galactic plane. A physical association between
Kwando and 47 Tuc can therefore be ruled out. Both E 3
and Pal 8 also lie within 1 kpc of Kwando's putative orbit, but
neither have published proper motions with which to
estimate their orbits.

\section{Conclusions} \label{conclusions}

A matched-filter examination of the southern Galactic cap in the PS-1
catalog reveals the existence of four new stellar stream candidates.
While Galactic substructures have been detected and mapped in the PS-1
survey area by others \citep{slater2014, morganson2016, bernard2016},
we attribute the discovery of these new stream candidates to a matched
filter reflecting a more metal-poor population, a different coordinate
projection, and perhaps a somewhat finer sampling in distance. Tenuous
streams such as these, with of order 10 stars per linear degree, are
easily swamped by the much larger number of foreground stars with
similar colors. Slight variations in filtering, sampling, projection,
foreground estimation, and general technique may be sufficient to
either highlight or possibly obscure such streams.  The results of
this investigation suggest that there may yet be a rich reservoir of
cold streams and substructures awaiting discovery in PS-1.

Based on their relatively narrow widths, and on memberships estimated
between $10^3$ and $10^4$ stars, we believe the progenitors of these
streams to have been globular clusters. They appear to occupy a
variety of orbits, from nearly circular to hyperbolic. While a number
of existing globular clusters lie near these trajectories, owing to
either divergent orbits or lack of proper motion measurements, we find
none that can be confidently associated with the new streams. We note
that many streams are have orbit normals that put them within the
region defining the Vast Polar Structure of
\citet{pawlowski2012}. Examination of Table 1 shows that both Molonglo
and Murrumbidgee appear to orbit within this plane.

We emphasize that the substructures identified here remain stream
candidates until they can be confirmed by radial velocity or proper
motion measurements.  This will presumably be enabled by the upcoming
second Gaia data release. Proper motion measurements in particular
will in most cases give us much better distances via Galactic parallax
\citep{eyre2009}, and ultimately enable more detailed modeling of the
Galactic potential.

\begin{acknowledgements}

We are very grateful to B. Shiao and R. White of the Space Telescope
Science Institute for their assistance in optimizing PS-1 database
queries. We also thank an anonymous referee for several useful
recommendations. 

The Pan-STARRS1 Surveys (PS1) have been made possible through
contributions of the Institute for Astronomy, the University of
Hawaii, the Pan-STARRS Project Office, the Max-Planck Society and its
participating institutes, the Max Planck Institute for Astronomy,
Heidelberg and the Max Planck Institute for Extraterrestrial Physics,
Garching, The Johns Hopkins University, Durham University, the
University of Edinburgh, Queen's University Belfast, the
Harvard-Smithsonian Center for Astrophysics, the Las Cumbres
Observatory Global Telescope Network Incorporated, the National
Central University of Taiwan, the Space Telescope Science Institute,
the National Aeronautics and Space Administration under Grant
No. NNX08AR22G issued through the Planetary Science Division of the
NASA Science Mission Directorate, the National Science Foundation
under Grant No. AST-1238877, the University of Maryland, and Eotvos
Lorand University (ELTE).

\end{acknowledgements}

{\it Facilities:} \facility{PS1}

\clearpage

\begin{figure}
\epsscale{1.0}
\plotone{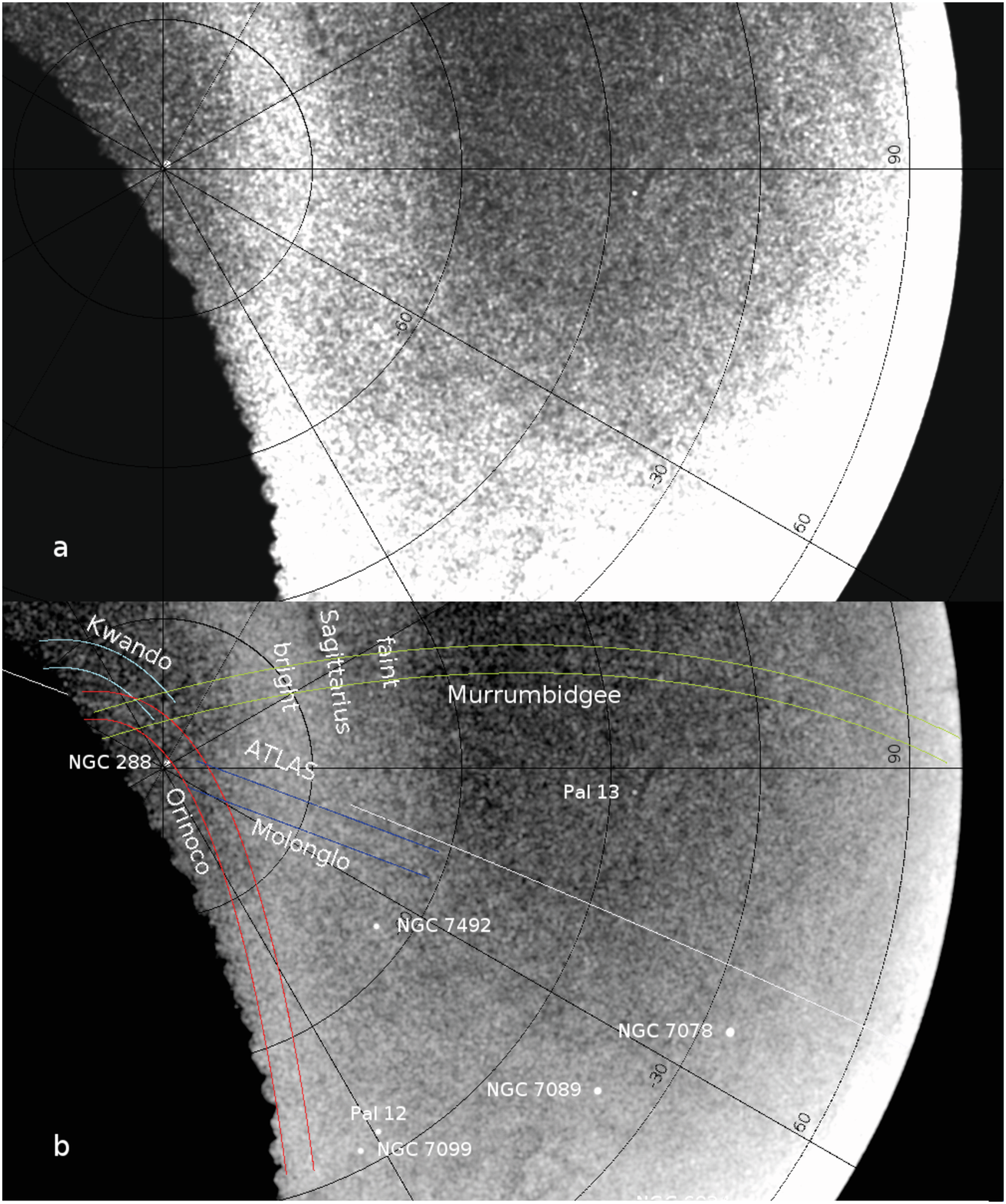}
\caption{Panel a: filtered surface density map of the PS-1 survey in
  the south Galactic cap, smoothed using a Gaussian kernel with
  $\sigma = 0.4\arcdeg$. The stretch is linear, with brighter areas
  indicating higher surface densities.  Panel b: The same surface
  density map with a log stretch, and with streams and nearby globular
  clusters labeled. The white lines show the great circle fit to the
  ATLAS stream by \citet{bernard2016}.}
\end{figure}

\begin{figure}
\epsscale{1.0}
\plotone{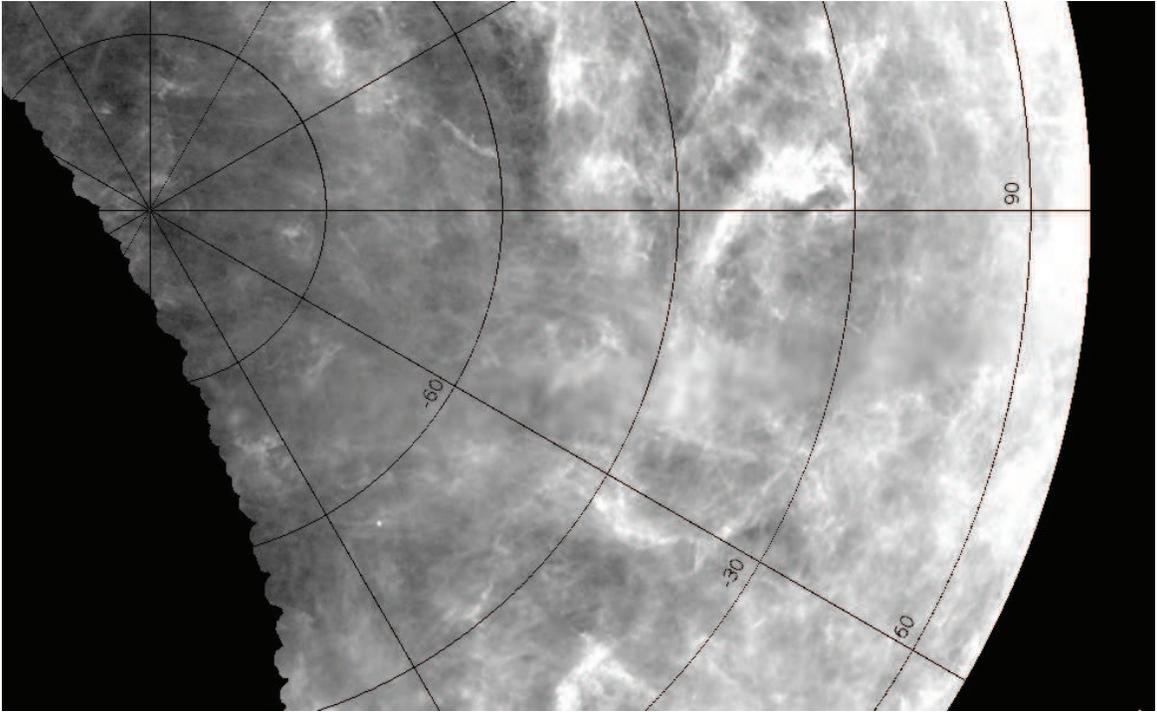}
\caption{The distribution of $E(B-V)$, as taken from \citet{schleg98},
  in the same coordinates as Figure 1.}
\end{figure}

\begin{figure}
\epsscale{1.0}
\plotone{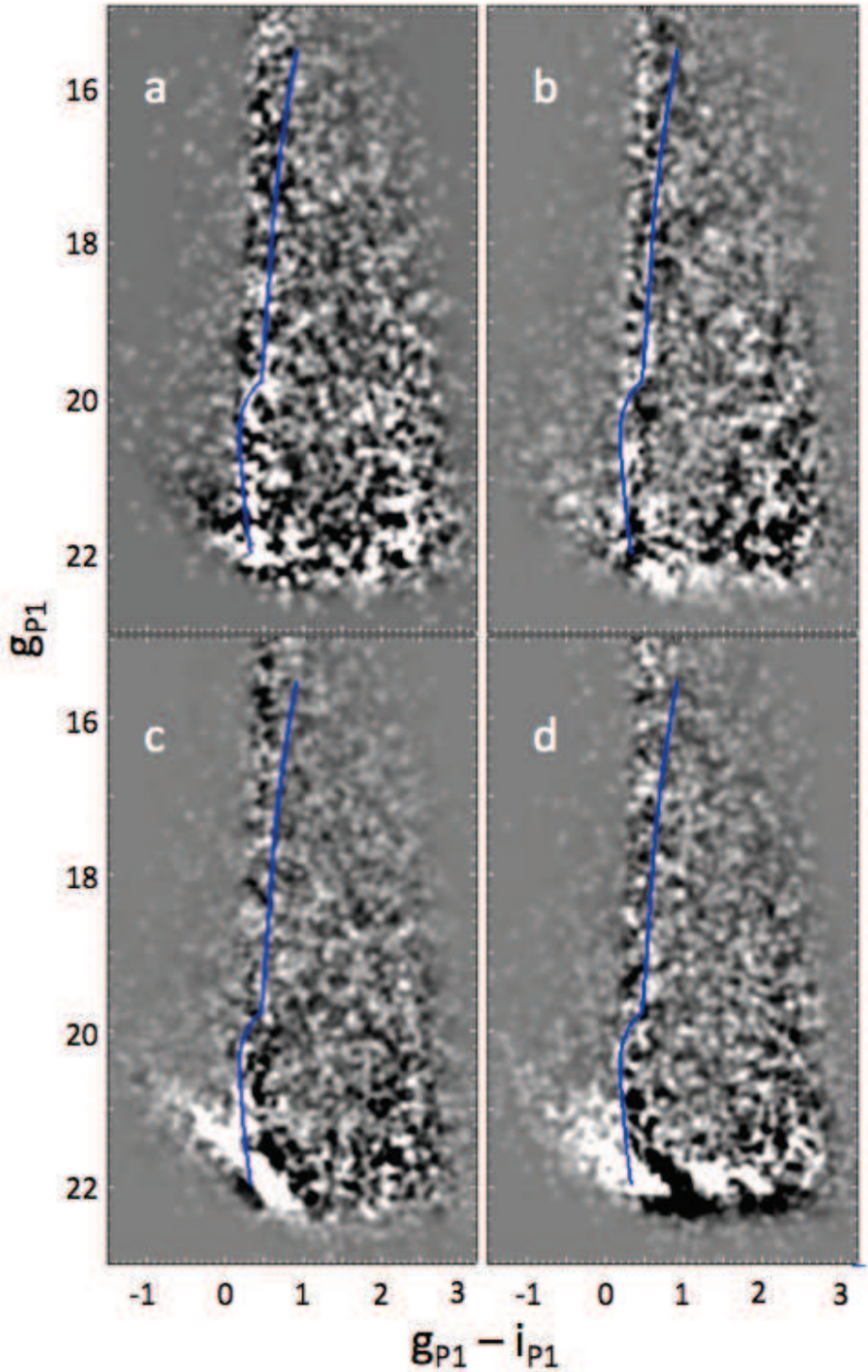}
\caption{Color-magnitude Hess diagrams of the (a) Molonglo, (b)
  Murrumbidgee, (c) Orinoco, and (d) Kwando. Lighter areas correspond
  to higher surface densities. The blue curve shows the
  main-sequence/giant branch locus measured for NGC 5053 and used as
  the basis of our matched filter, shifted to a distance of 20
  kpc. The CMDs of the streams were estimated by generating the CMDs
  of all stars lying within $0.25\arcdeg$ of each stream and
  subtracting from these the normalized distributions of stars in
  selected, much larger fields surrounding each stream. For
  Murrumbidgee the CMD was sampled over the region $-65\arcdeg < b <
  -30\arcdeg$, while for Orinoco we used the eastern $19\arcdeg$ of
  the stream. For Molonglo and Kwando we used the full lengths of the
  streams as given in the text.}
\end{figure}

\begin{figure}
\epsscale{1.5}
\plotone{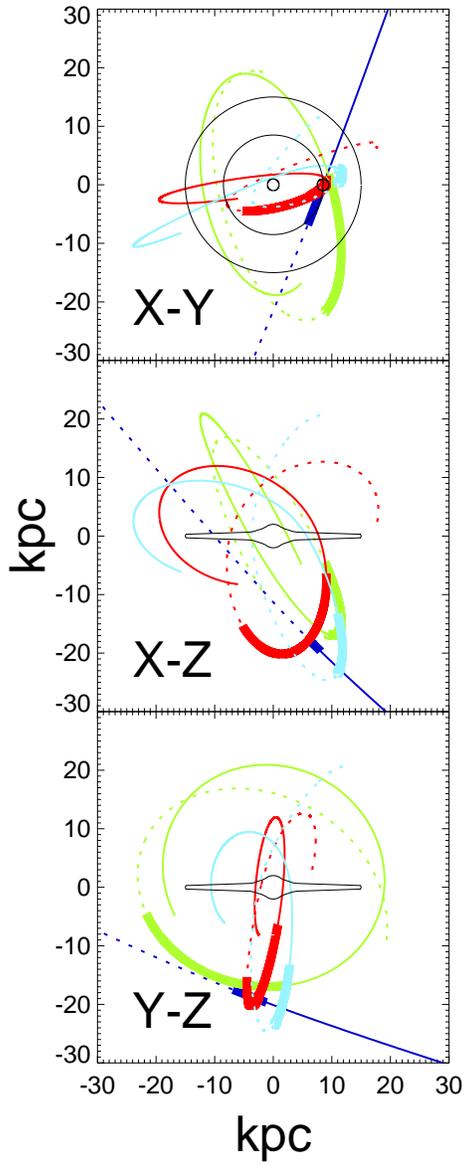}
\caption{Integrated best-fit orbits, in Galactic cartesian
  coordinates, for the streams in Figure 1. The position of the sun
  and the solar circle are shown in the uppermost panel. The colors
  are the same as those used in Figure 1. The dotted lines show the
  forward integrations of the prograde orbits, while the solid lines
  show the rearward integrations. The thick portions of the orbits are
  the regions for which we have PS-1 data.}
\end{figure}

\begin{deluxetable}{clccccc}
\tabletypesize{\small}
\tablecaption{Predicted Motions and Orbit Parameters}
\tablecolumns{5}

\tablehead{& & \colhead{Molonglo} & \colhead{Murrumbidgee} & \colhead{Orinoco} & \colhead{Kwando}} \\
\startdata
Fiducial Point & R.A. ($^\circ$, J2000) & 0.163 & 358.614 & 359.701 & 25.025 \\
\\
& dec ($^\circ$, J2000) & -16.869 & +16.274 & -25.324 & -25.107 \\
\hline \\
& $v_{hel}$ (km s$^{-1}$) & $+6.1^{+73}_{-329} $ & $-123 \pm 50$ & $+64 \pm 3.5$ & $ +130 \pm 35 $ \\
Prograde Orbit & $\mu_\alpha$ cos($\delta$) (mas yr$^{-1}$) & $+8.08 \pm 0.13$ & $-0.134 \pm 0.009$ & $+0.253 \pm 0.002$ & $+0.969 \pm 0.004$ \\
& $\mu_\delta$ (mas yr$^{-1}$) & $-8.15 \pm 0.13$ & $+0.388 \pm 0.01$ & $-2.28 \pm 0.0012$ & $ -1.633 \pm 0.003$ \\ 
\\
\hline \\
& $v_{hel}$ (km s$^{-1}$) &  $-118^{+330}_{-73}$ & $-182 \pm 50$ & $ -121 \pm 3.5 $& $ -94 \pm 35$ \\
Retrograde Orbit & $\mu_\alpha$ cos($\delta$) (mas yr$^{-1}$) & $-5.91 \pm 0.13$ & $+2.152 \pm 0.009$ & $+1.91 \pm 0.002$ & $+1.865 \pm 0.004$ \\
& $\mu_\delta$ (mas yr$^{-1}$) & $+3.96 \pm 0.12 $ & $-3.074 \pm 0.011$ & $-2.06 \pm 0.001 $ & $-2.009 \pm 0.003$ \\
\\
\hline \\
$R_{apo}$ (kpc) && $> 100$ & $ 24^{+4}_{-1}$ & $21 \pm 0.15 $ & $ 26.4^{+3.2}_{-1.8}$\\
$R_{peri}$ (kpc) && $20\pm 0.2$ & $21^{+0.1}_{-2.2}$ & $6.9 \pm 0.1$ & $ 4.4 \pm 0.7$\\
$i$ ($^\circ$) && $68 \pm 1$ & $62 \pm 4$ & $79.7 \pm 0.1$ & $ 76.7 \pm 4$ \\
$\epsilon$ && $ 1 \pm 0.3$ & $0.08^{+0.13}_{-0.001}$ & $ 0.5 \pm 0.003$ & $ 0.72 \pm 0.02$\\
Orbit Pole & $l$ ($^\circ$) & $193^{+9}_{-2}$ & $160 \pm 4 $& $ 250 \pm 0.3$ & $ 237 \pm 15$ \\
&            $b$ ($^\circ$) & $+22 \pm 1$ & $+28 \pm 4 $& $+10.3 \pm 0.1 $ & $ +13 \pm 4$ \\
\enddata
\end{deluxetable}

\end{document}